# Using Nudges to Prevent Student Dropouts in the Pandemic

Guilherme Lichand[1]* & Julien Christen[2]*

**The impacts of COVID-19 reach far beyond the hundreds of thousands of lives lost to the disease; in particular, the pre-existing learning crisis is expected to be magnified during school shutdown. Despite efforts to put distance learning strategies in place, the threat of student dropouts, especially among adolescents, looms as a major concern. Are interventions to motivate adolescents to stay in school effective amidst the pandemic? Here we show that, in Brazil, nudges via text messages to high-school students, to motivate them to stay engaged with school activities, substantially reduced dropouts during school shutdown, and greatly increased their motivation to go back to school when classes resume. While such nudges had been shown to decrease dropouts during normal times, it is surprising that those impacts replicate in the absence of regular classes because their effects are typically mediated by teachers (whose effort in the classroom changes in response to the nudges). Results show that insights from the science of adolescent psychology can be leveraged to shift developmental trajectories at a critical juncture. They also qualify those insights: effects increase with exposure and gradually fade out once communication stops, providing novel evidence that motivational interventions work by redirecting adolescents' attention.**

[1] Department of Economics, University of Zurich, Zurich, Switzerland. [2] Center for Child Well-Being and Development, University of Zurich, Zurich, Switzerland. * e-mail: guilherme.lichand@econ.uzh.ch; julien.christen@econ.uzh.ch

The COVID-19 pandemic has forced over 1.5 billion children and adolescents across 160 countries to stay at home while schools were shut down on sanitary grounds[1]. Research on the impacts of similar historical events, such as the Spanish flu, Ebola and the H1N1 epidemics, has documented that pandemics not only cost lives and employment, but also substantially deteriorate learning outcomes of school-age children[2,3,4], with persistent impacts on their future labor market outcomes[2]. Multiple forces combine to push children out of school in such settings: lower returns to education in face of the economic crunch, demand for child labor among poor families, violence against children in a context of stress and with children at home, and loss of motivation to go to school in the absence of face-to-face interactions with teachers and peers[5,6,7,8,9,10,11]. As such, a surge in student dropouts is expected to take place in 2020 unless drastic actions are put in place[12,13].

While significant attention has been devoted to interventions that can support distance learning, particularly to mitigate learning deficits when children come back to school[14,15,16,17], a large fraction of children and adolescents might actually never go back[1]. That concern is particularly acute among middle- and high-school students, inspired by insights from a large literature on adolescent psychology. This literature documents that neuro-biological changes during puberty redirect adolescents' attention and motivational salience[18], with status-seeking behaviors, romantic interests and peer pressure[18] often getting in the way of attending classes. As a result, adolescents are the ones most likely to drop out in normal times[6], and presumably even more in the context of the COVID-19 crisis.

If insights from adolescent psychology suggest that this population is the one most at risk of dropping out amidst the pandemic, they also lay out opportunities to intervene. In fact, motivational interventions have been shown to successfully improve adolescents' choices, from healthy eating[19] to school effort[20], with the potential to shift developmental trajectories. Could motivational interventions also be used effectively to encourage adolescents to stay in school amidst the pandemic, particularly in poor countries, where pressures to drop out of school might be amplified?

Recent evidence shows that, in developing countries, nudges to motivate public school students to engage with school activities have the potential to not only significantly improve learning outcomes[21,22], but also to drastically decrease dropouts during normal times[23]. Having said that, it would be surprising if those effects replicated in the absence of regular classes, because they are typically mediated by teachers, whose effort in the classroom changes in response to the nudges[23].

Here we show that an intervention to motivate high-school students to stay engaged with school activities during the pandemic substantially prevented student dropouts in Brazil *even in the absence of regular classes*.

**Research Design and Intervention**

To study this question, we undertook a Randomized Control Trial in the State of Goiás, Brazil (pre-registered as trial 5986 at the AEA RCT Registry), in partnership with Instituto Sonho Grande and the Goiás State Secretariat of Education in the context of their full-time high school program, "Ensino Médio em Tempo Integral". In Goiás, face-to-face classes were suspended in March 2020, and are not expected to resume until October (and even then, only if new COVID-19 cases in the State are kept under control). During the school shutdown, classes switched to online, delivered through a video conferencing and team collaboration platform. Students were assigned daily exercises that they had to hand in through the platform. For those without internet access, schools handed out assignments in plastic bags hung at the school front door, and students had to hand them back in the same way.

The intervention, powered by Movva (the implementing partner of Eduq+, the educational nudgebot evaluated in the interventions mentioned above[21,23]), consisted of sending nudges twice a week over text messages (SMS) to high-school students or their primary caregivers. Whenever there were multiple phone numbers on record for a student, we randomized which would be targeted by the intervention. Nudges comprised encouragement messages meant to have students engage in distance learning activities (online and offline) and to keep them motivated about staying enrolled in school when face-to-face classes resume. In total, 12,056 high-school students across 57 public schools received nudges between June 9th and July 3rd, while other 6,200 high-school students across 30 public schools received no nudges or other SMS from their schools. The intervention ended after roughly one month because of the winter break.

**Definition of Outcomes and Estimation**

In order to evaluate the effectiveness of the intervention, we monitored (1) *student dropouts during school shutdown*, based on administrative records shared by the Education Secretariat, equal to 1 if a student has no attendance on record for two weeks in a row right before the winter break, and 0 otherwise; and (2) *lack of motivation to return to school once they reopen*, based on students' self-reports, equal to 1 if a student states no intention of going back to school once regular classes resume, and 0 otherwise. We elicited the latter over SMS, from rotating sub-samples of students in the treatment and control groups every week – from the week after the intervention started until 3 weeks after it ended.

We estimate treatment effects on those two outcomes, and leverage weekly self-reported data to study how effects vary with exposure to the nudges, and the extent to which they persist or fade-out after communication ceases. Since we cannot verify whether students effectively received messages as intended, all estimates presented are intention-to-treat (ITT) analyses.

While the Secretariat had information on whether phone numbers belonged to the students or to their primary caregivers, we found out ex-post that records were often inaccurate and that handsets were often shared within the household; as such, we do not explore heterogeneous treatment effects by who was targeted by the intervention. Since there is no data on learning outcomes (as standardized tests are not planned for the remainder of the school year, given the exceptional circumstances), we cannot evaluate treatment effects on student proficiency.

**Effects on Student Dropouts during School Shutdown and Motivation to Return to School**

Figure 1 displays average treatment effects on student dropouts during school shutdown (based on administrative data, in Panel A) and on their lack of motivation to return to school once they reopen (using self-reported data only while the intervention lasted, in Panel B). Panel A showcases that while nearly 6% of students in the control group had dropped out of distance learning activities right before the winter break, that figure was only 1.4% in the treatment group – a 77.3% reduction (p-value = 0.006). Next, Panel B documents that while nearly 24% of students in the control group lacked motivation to go back to school when regular classes resume, that figure was about 13.5% in the treatment group – a 43.7% reduction (p-value = 0.000).

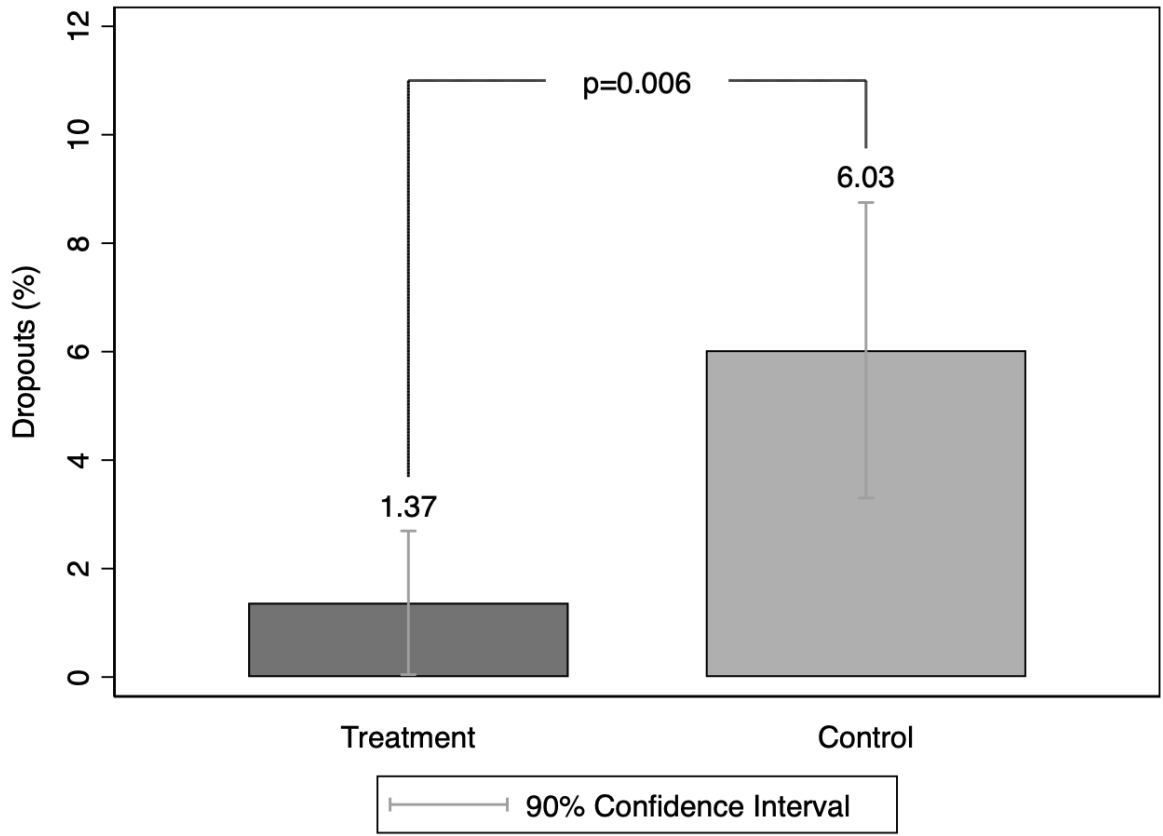

**Figure 1 – Panel A:** Treatment effects of SMS nudges on student dropouts during school shutdown (based on administrative data)

> **Notes:** ITT estimate from an Ordinary Least Squares (OLS) regression with dropouts = 1 if a student had no attendance on record over the last two weeks before the winter break, and 0 otherwise. 90% confidence intervals in light grey brackets; p-value in dark grey brackets from a test of equality of proportions between the treatment and control groups, with standard errors clustered at the school level. In the Supplementary Materials, Table S1 shows that student characteristics are balanced across the treatment and control groups; Table S4 documents that results are robust to defining dropouts alternatively as no attendance on record over the last week or over the last three weeks before the winter break.

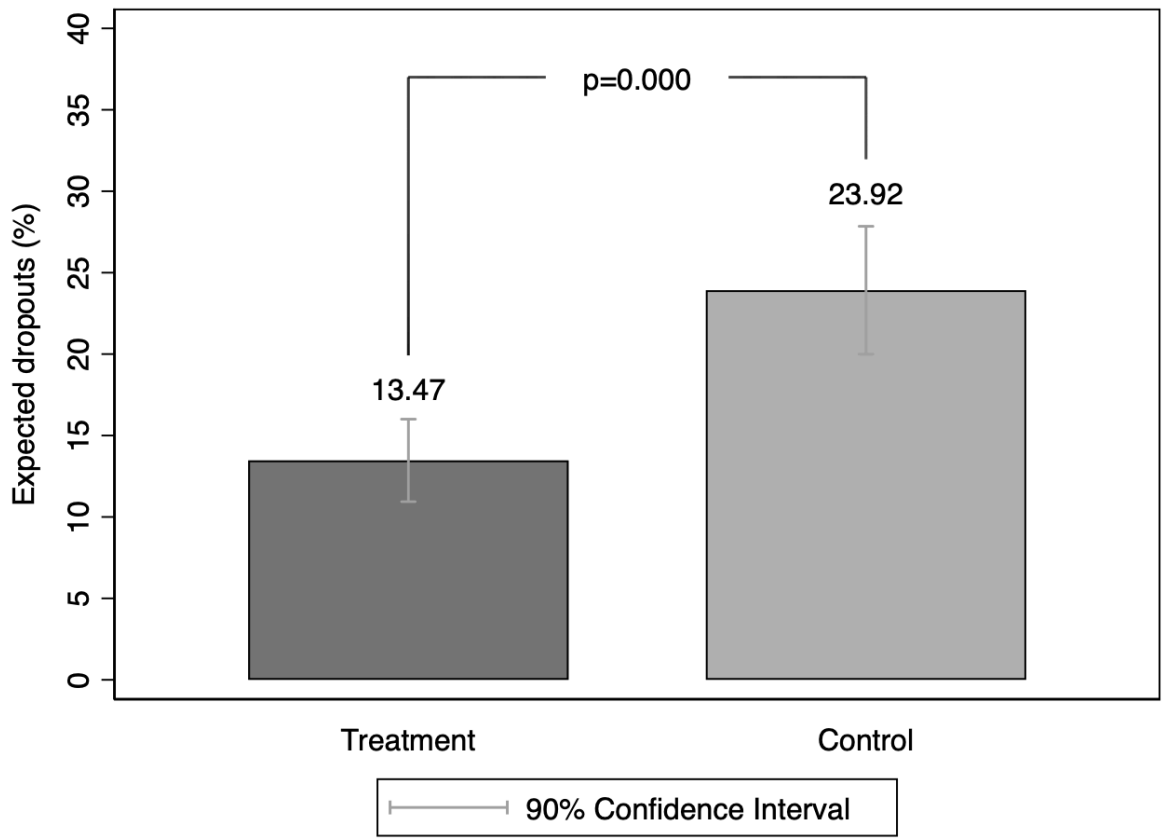

**Figure 1 – Panel B:** Treatment effects of SMS nudges on students' lack of motivation to return to school once they reopen (based on self-reported data)

> **Notes:** ITT estimate from an Ordinary Least Squares (OLS) regression with expected dropouts = 1 if the student states that s/he does not think s/he will be back in school when regular classes resume, and 0 otherwise. 90% confidence intervals in light grey brackets; p-value in dark grey brackets from a test of equality of proportions between the treatment and control groups, with standard errors clustered at the school level. In the Supplementary Materials, Table S1 shows that student characteristics are balanced across the treatment and control groups; Table S2 shows that the probability of responding to SMS surveys is not systematically affected by the treatment.

**Attention or Information? The Onset and Persistence of the Effects of the Intervention**

While the intervention is designed with the goal of *redirecting adolescents' attention* to the benefits of remaining engaged with school activities amidst the pandemic, it could be that its effects on motivation and behavior work through alternative mechanisms. For instance, it could be that messages make students believe that their teachers care more about them than they originally thought.

We take advantage of our unique data on students' motivation, available at a weekly basis, to study this question. We do so by estimating how quickly the effects of the intervention kick in after it starts, and the extent to which they persist after communication ceases. If the intervention works by inducing students to update beliefs based on information they infer from school communication (as in the example of students being surprised about how much teachers seem to care about them), motivation should respond immediately to the nudges, and their effects should persist even after communication ceases. In contrast, if the intervention works by redirecting students' attention, then motivation should gradually build up after the intervention starts, and gradually die out once it stops.

     Figure 2 displays the prevalence of students' lack of motivation to return to school, week by week, across the treatment and control groups (Panel A), and week-by-week treatment effects (Panel B). Panel A documents a striking pattern for expected dropouts in the control group, which increased by nearly 2-fold in little over a month (starting from 15% by the 2$^{nd}$ week of June and reaching nearly 40% by the 3$^{rd}$ week of July). Panel A also shows that expected dropouts not only started from a lower level in the treatment group already by week 2, but also increased at slower rates while the intervention lasted. Panel B confirms those patterns: nudges decreased expected dropouts by over 30% already by week 2, and effect sizes increased with the length of exposure, reaching a 50% reduction by week 4 (significant at the 1% level).

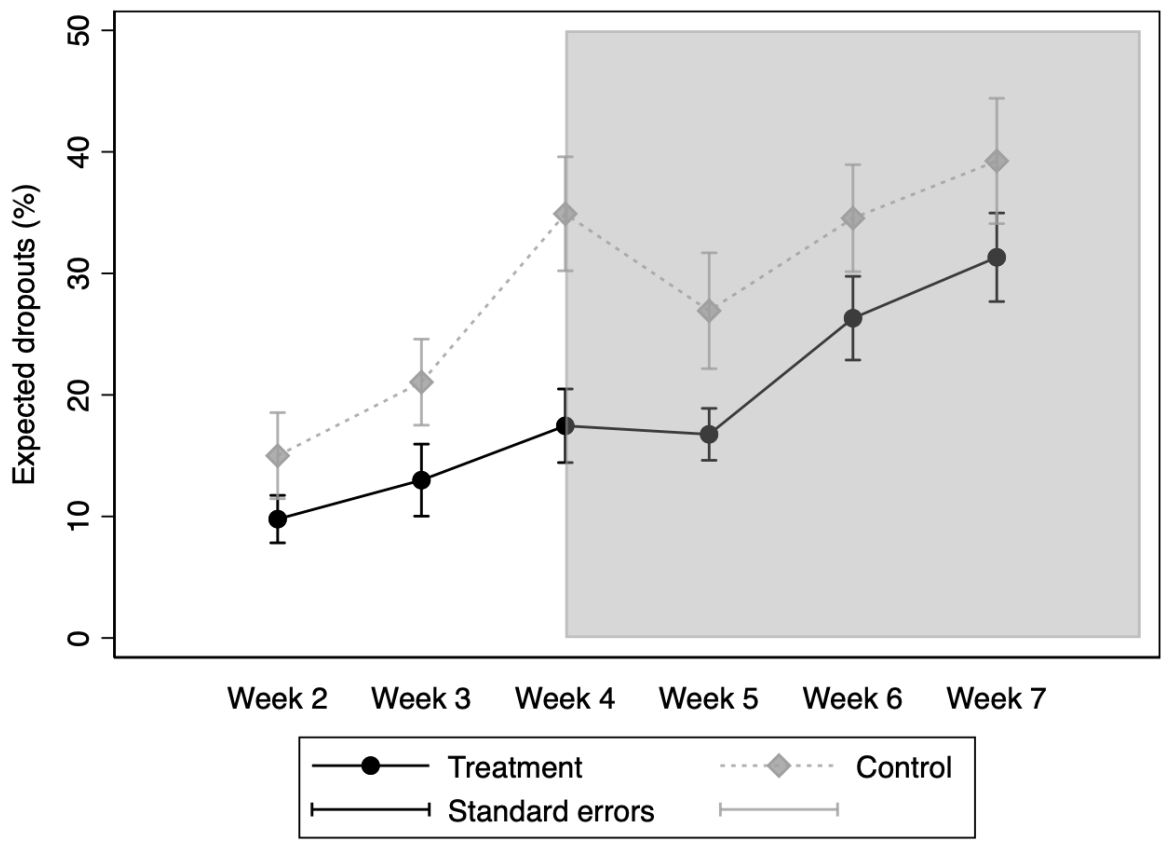

**Figure 2 – Panel A:** Students' lack of motivation to return to school once they reopen (based on self-reported data) for the treatment and control groups, week by week

**Notes:** Weekly sample averages for expected dropouts (= 1 if the student states that s/he does not think s/he will be back in school when regular classes resume, and 0 otherwise) for the treatment group (in black) and the control group (in light grey). Self-reports based on SMS surveys from rotating sub-samples of students in the treatment and control groups every week – from the week after the intervention started until 3 weeks after it ended. Standard errors clustered at the school level. The shaded area corresponds to the weeks after communication ceased.

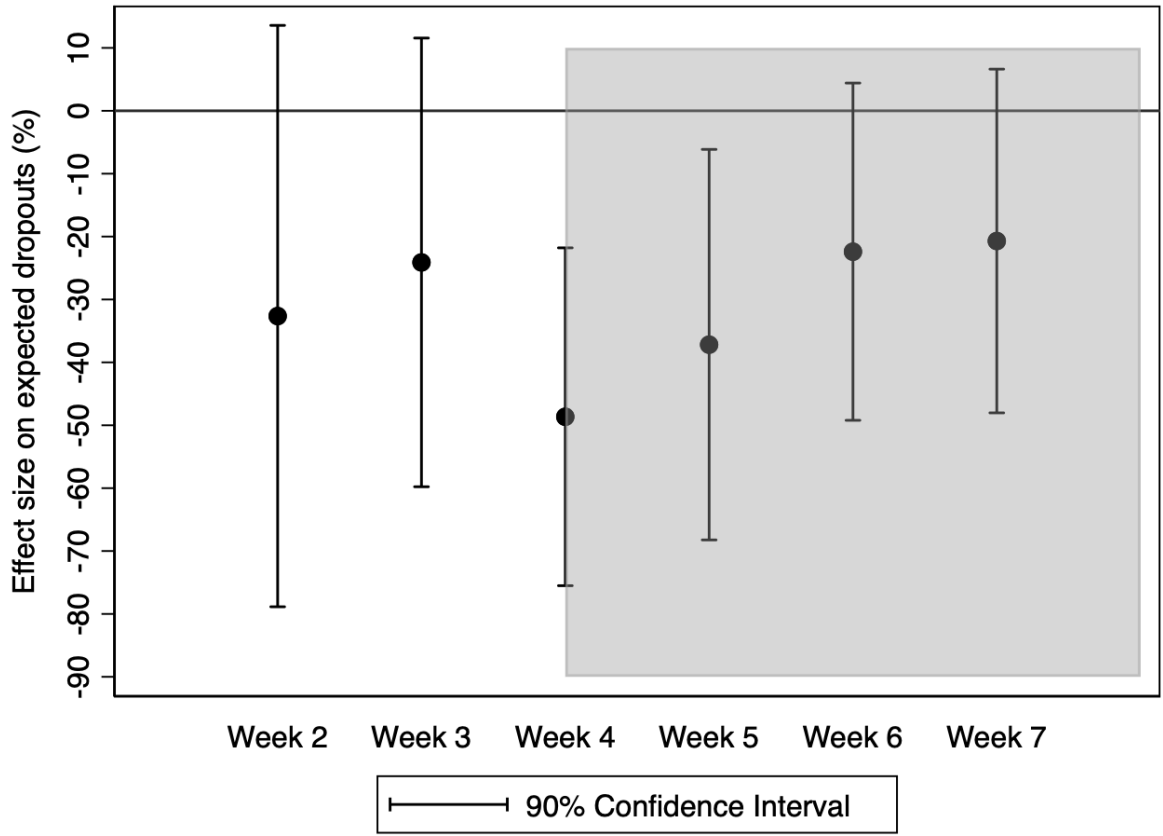

**Figure 2 – Panel B:** Treatment effects of SMS nudges on students' lack of motivation to return to school once they reopen (based on self-reported data), week by week

**Notes:** ITT estimates from Ordinary Least Squares (OLS) regressions week by week with expected dropouts = 1 if the student states that s/he does not think s/he will be back in school when regular classes resume, and 0 otherwise. 90% confidence intervals in dark grey brackets, with standard errors clustered at the school level. The shaded area corresponds to the weeks after communication ceased. In the Supplementary Materials, Table S1 shows that student characteristics are balanced across the treatment and control groups for all weeks except week 3; for this reason, we control for student characteristics in all regressions week by week. Table S3 shows that the probability of responding to SMS surveys is not systematically affected by the treatment at any week.

After the intervention ended, treatment effects gradually receded. Panel A in Figure 2 shows that expected dropouts increased nearly by the same rate across treatment and control groups after communication ceased. Panel B documents that, while treatment effects were still large and statistically significant (at the 5% level) one week after the end of the intervention, effect sizes quickly declined, reaching 20% two weeks later, no longer statistically different from zero.

**Discussion**

Results document that simple motivational interventions to support and engage adolescents while schools are shut down – largely overlooked by policy-makers amidst efforts to keep curricular activities running in the absence of face-to-face classes – can effectively prevent student dropouts. The intervention that we evaluate could be implemented at scale almost anywhere, as cell phone penetration is very high worldwide[24], and as SMS does not require smartphones or internet access. Even in face of illiteracy challenges, there is recent evidence that nudges over text messages can work just as well as audio messages[23]. Our findings provide important lessons to address the global education crisis in the context of the pandemic: above and beyond focusing on curricular knowledge to address learning deficits, public school systems should reach out to families to provide support and encouragement during challenging times.

The patterns of how effect sizes vary with exposure to the treatment, gradually building up as the nudges continue and gradually fading out once they stop, provide novel evidence connecting the effects of motivational interventions to findings from the literature on adolescent psychology. While seminal work notes that the influence of pubertal maturation on neuro-behavioral development leads to a re-orientation of adolescents' attention and motivational salience, studies evaluating motivational interventions in this context typically cannot distinguish whether they work because they redirect subjects' attention to desirable goals or because of alternative mechanisms (e.g. because they induce subjects to update beliefs based on what they infer from the intervention). Our results showcase first-hand that, as with adult populations, and as in other decision domains[25,26,27], motivational nudges to keep adolescents in school work by affecting what issues are *top-of-mind* and changing behavior accordingly.

That distinction matters not only for understanding the scientific underpinnings of behavior change, but also for policy design. This result suggests that if, on the one hand, quickly rolling out nimble interventions to engage public school families amidst the pandemic could allow for quick wins, on the other hand, once those interventions are discontinued, their impacts are likely to gradually die out. As such, public school systems would be better-off setting up institutional programs to support families' and students' engagement over a longer time frame, rather than putting in place short-lived interventions that cannot be embedded in their regular activities.

**METHODS**

**Ethics Approval.** Approval for this study was obtained from the Institutional Review Board of the Department of Economics at the University of Zurich (2020-033). This experiment was conducted in the context of the full-time high school program "Ensino Médio em Tempo Integral" of the Goiás State Secretariat of Education. When it comes to informed consent, since participants are minors, broad consent was obtained from their legal guardians directly by the Education Secretariat (at the time of school enrollment), allowing researchers to use secondary information from administrative records without eliciting further consent. Our implementing partner Movva further obtained students' assent directly via text messages (SMS): participants were informed and reminded of the fact that they could opt-out from the intervention and SMS surveys at any point (by simply replying 'STOP' or 'CANCEL', free of charge), without consequence.

**Participants.** Participants consist of public school students enrolled in grades 10-12; typical age is 15-18 years old. All contacts were provided to Movva by the Goiás State Secretariat of Education. The total number of contacts in the database correspond to 18,256 students, 12,056 of which randomly assigned, across 57 schools, to receive SMS nudges between June 9$^{th}$ and July 3$^{rd}$, and 6,200 across 30 schools assigned not to receive nudges or any other SMS communication from their schools. Power calculations before the onset of the intervention pointed out this sample size was large enough to detect relevant minimum effects on the outcomes of interest.

**Data collection.** Before the start of the intervention, the contacts' database was shared with the authors to complete the randomization at the school level, stratified by gender, grade and phone ownership. Schools were randomly assigned to either a treatment or a control group, following the group sizes above and using the statistical software Stata. The database including a treatment assignment indicator was then returned to Movva such that SMS nudges could be sent accordingly to the treatment group, but not to the control group.

Data on online access to the platform and participation in offline school activities was shared by the Secretariat of Education with Movva, while data on motivation to return to regular classes once they resume was collected by Movva directly over SMS surveys, from rotating sub-samples of approximately 280 students in the treatment and control groups every week – from the week after the intervention started until 3 weeks after it ended. Weekly sub-samples were also randomly drawn from the subject pool. Balance tests using Wald tests of simple and composite linear hypotheses were conducted to ensure that each treatment group is comparable with respect to students' characteristics that are gender, grade in which the student is enrolled, and individual phone ownership. These tests and results are detailed in the supplementary material to this paper in Table S1.

Outcome data was shared weekly with the authors, and analyzed following a pre-analysis plan pre-registered as [trial 5986](#) at the AEA RCT Registry (included as supplementary material to this paper). All analyses were conducted by the authors using the statistical software Stata.

Finally, collecting information on human participants over time is subject to attrition. Participants were free to leave the study at any time, which creates a risk of biasing the results if such attrition is correlated with treatment assignment. In this context, we tested whether the probability of students responding to the SMS surveys was affected by the treatment. The results, which are reported in the supplementary material to this paper in Tables S2 and S3, indicate that the probability of responding to SMS surveys is not systematically affected by the treatment.

**Intervention.** Movva, the start-up that powered the intervention evaluated in this study, specializes in promoting behavior change by sending frequent reminders and encouragement messages directly to users' cell phones. The concept of nudges – interventions that modify the choice architecture by changing the way decisions are framed to mitigate or amplify behavioral biases, inducing certain decisions while preserving subjects' freedom of choice – lies at the heart of the contributions of Nobel Memorial Prize in Economic Sciences winner Richard Thaler and co-authors. Nudges have been shown to effectively change behaviors across various contexts, from education to preventive health care to savings[25]. Eduq+, the intervention evaluated in this study, has been shown to improve educational outcomes in an environment of regular classes across different settings[21,23]. In the context of this study, two nudges per week were sent over text messages (SMS) to high-school students or their primary caregivers, depending on phone ownership, in the treatment group. Nudges were organized in 2-week sequences of 4 messages, as follows (translated from Portuguese):

- Sequence 1
    - Fact: « EDUQ+: It is normal to be afraid in times of uncertainty. Use this scenario to your advantage: take the opportunity to develop the ability to focus on your plans for the future. »
    - Activity: « EDUQ+: How about summarizing your life project? Highlight which dreams you would regret NOT realizing. Plan step by step how to get there. »
    - Interactivity: « EDUQ+: Tell us! From 0 to 10, what is your level of confidence that completing high school will help with your plans for the future? SMS free of charge. »
    - Growth: « EDUQ+: One step at a time! That's how we build our story. Be the protagonist of yours and focus on your studies to finish the school year. »
- Sequence 2
    - Fact: « EDUQ+: Connect! 80% of your colleagues believe in high school to help them do well in the future. To get there, you need to be a friend of the clock when studying! »
    - Activity: « EDUQ+: Time to study and marathon! Make a schedule of the day, setting time to wake up, study, do activities and, of course, catch up on video lessons. »
    - Interactivity: « EDUQ+: A day used to study gets you closer to the degree. How has your time management been? 1. Good 2. Regular 3. Bad. SMS free of charge. »
    - Growth: « EDUQ+: Deserved holidays! Time to rest, but without losing focus on the future. Between one leisure activity and another, pay attention to what you like and helps with learning. »

**Measures.** *Student dropouts during school shutdown*. The Secretariat of Education shared administrative records with Movva, which then shared it with the authors. This dataset contained information on access to the online platform and participation in offline school activities at the student level, an indicator variable equal to 0 if the student logged in to the platform or participated in offline school activities on a given day and 1 in case she/he did not. Based on this information, we created a measure of dropouts which was equal to 1 if a student had no attendance on record for two weeks in a row right before the winter break (between June 15$^{th}$ and June 26$^{th}$), and 0 otherwise. To ensure that the results presented in the paper are robust to alternative definitions of dropouts, we constructed analogous measures of dropouts as no attendance on record for the last week before holidays or for the last three weeks before holidays. Results, which are very robust to using alternative definitions of dropouts, are reported as supplementary material to this paper, in Table S4.

*Motivation to return to school once they reopen*. Each week, students assigned to be surveyed by text message reported their motivation to return to school once regular classes resume by answering the following question: "Do you plan on returning to school once regular classes resume?". Movva coded lack of motivation to return as a binary indicator based on SMS replies, equal to 1 if the reply was "No" or similar, and zero otherwise, and shared that information with the authors.

**Analysis method.** All results presented in the paper use intention-to-treat analyses by linking student identification numbers to the treatment condition they were assigned to before the start of the intervention. The reason for restricting our analyses to intention-to-treat analyses is that we had no means of verifying whether students effectively received messages as intended. Throughout the paper, we report intention-to-treat effects obtained from Ordinary Least Squares (OLS) regressions, by regressing each outcome on a binary indicator equal to 1 if the student was assigned to treatment and 0 otherwise. In Panel B of Figure 2, effect sizes are standardized by dividing the treatment effect coefficients by the standard deviation in the control group for this outcome, providing effect sizes in % terms. Last, since for one of the weeks of our SMS surveys, student characteristics were not balanced across respondents in the treatment and control groups, we control for gender, grade in which the student is enrolled and phone ownership in this figure. All p-values are obtained from two-tailed tests of equality of proportions between the treatment and control groups, with standard errors clustered at the school level in each case.

**Reporting summary.** Further information on research design is available in the Research Reporting Summary linked to this paper.

**Data Availability**
The data that support the findings of this study are available at https://osf.io/3sqfr/ or upon reasonable request to the authors.

**Code Availability**
Syntax can be found at https://osf.io/3sqfr/ or upon reasonable request to the authors.

**Acknowledgements**
This manuscript uses administrative data on students' attendance, granted by the Goiás Secretariat of Education. Funding towards the intervention evaluated in this study was provided by Instituto Sonho Grande. We acknowledge helpful comments during the preparation of the manuscript from Ernst Fehr, Johannes Haushofer, Brad Wible and David Yeager. The content is solely the responsibility of the authors.

**Author contributions**
G.L. conceived the study and led the design, analysis and writing; J.C. was involved in every phase of the study, particularly the conception of the study, the study design, the interpretation of analyses and the writing of the manuscript.

**Author information**
G.L. is a co-founder and chairman at Movva, the implementing partner of the intervention evaluated in this study.

**Supplementary Materials**

**Figure S1** – Histogram of data availability for students' attendance based on administrative records

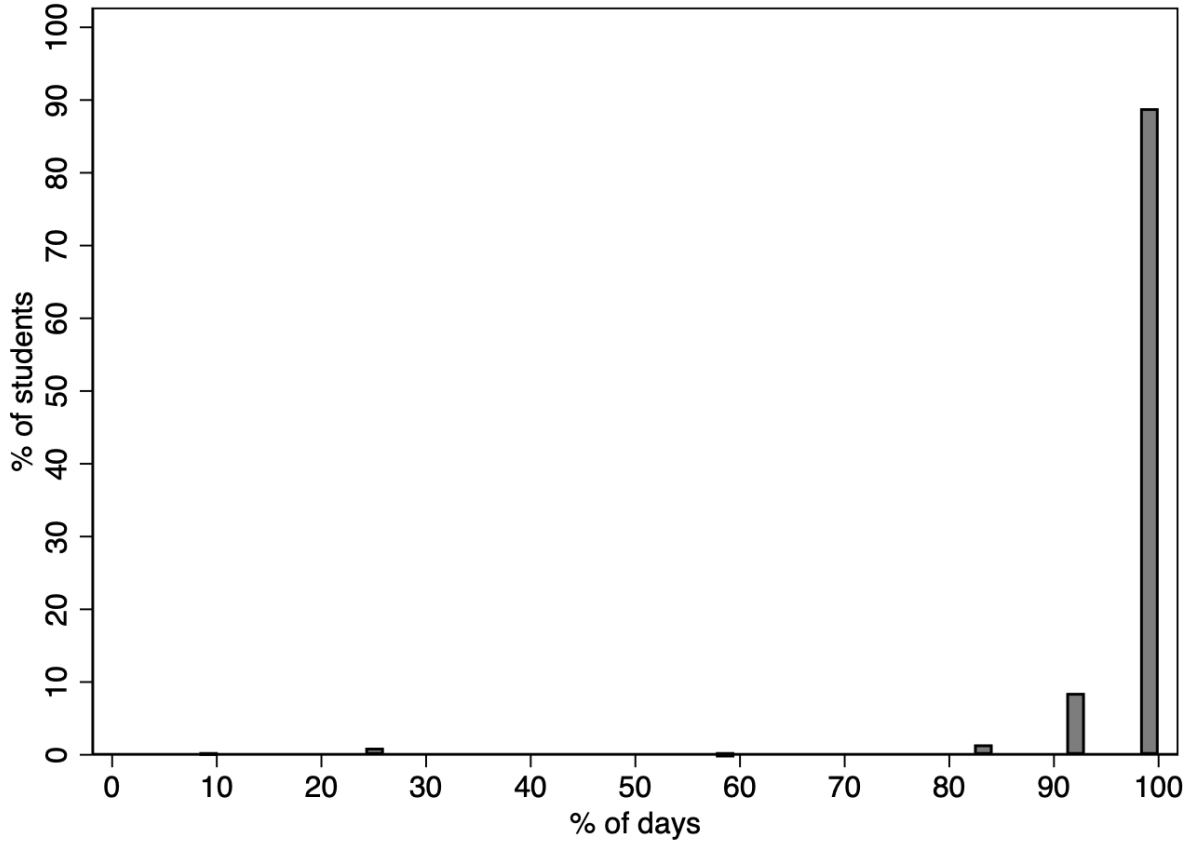

**Notes:** Histogram of data availability for students' attendance based on administrative records, for the school days after the start of the intervention (June 9-26; except June 11-12, which were school holidays). The distribution is left-skewed, with full records available for nearly 90% of students. The % of days is computed by dividing the number of observations for each student after the start of the intervention by 12 and multiplying by 100.

## Table S1 – Balance tests
### Panel A – All students

|  | Sub-sample means | | Diff.=0 [p-value] | Number of observations |
|---|---|---|---|---|
|  | Treatment | Control |  |  |
| **Variables** |  |  |  |  |
| Gender | 0.46 | 0.48 | 0.197 | 18,256 |
| Grade | 1.83 | 1.84 | 0.710 | 18,256 |
| Student owns phone | 0.43 | 0.41 | 0.690 | 18,256 |
| F-test |  |  | 0.515 | 18,256 |

**Notes:** Ordinary Least Squares (OLS) regressions of student characteristics on treatment assignment at the time of randomization. P-values from a test of equality of means of the treatment and control groups for each covariate (jointly for the F-test), with standard errors clustered at the school level. Results indicate that student characteristics are balanced across the treatment and control groups.

### Panel B – Students with valid outcome data (used in regressions)

|  | Sub-sample means | | Diff.=0 [p-value] | Number of observations |
|---|---|---|---|---|
|  | Treatment | Control |  |  |
| **Administrative data** |  |  |  |  |
| Gender | 0.46 | 0.48 | 0.192 | 18,157 |
| Grade | 1.83 | 1.84 | 0.706 | 18,157 |
| Student owns phone | 0.42 | 0.41 | 0.661 | 18,157 |
| F-test |  |  | 0.501 | 18,157 |
| **Self-reported data (weeks 2 to 4)** |  |  |  |  |
| Gender | 0.45 | 0.40 | 0.243 | 828 |
| Grade | 1.81 | 1.84 | 0.637 | 828 |
| Student owns phone | 0.20 | 0.23 | 0.481 | 828 |
| F-test |  |  | 0.505 | 828 |
| **Self-reported data – week 2** |  |  |  |  |
| Gender | 0.43 | 0.40 | 0.645 | 284 |
| Grade | 1.74 | 1.95 | 0.074 | 284 |
| Student owns phone | 0.17 | 0.16 | 0.845 | 284 |
| F-test |  |  | 0.387 | 284 |
| **Self-reported data – week 3** |  |  |  |  |
| Gender | 0.44 | 0.39 | 0.377 | 249 |
| Grade | 1.88 | 1.78 | 0.311 | 249 |
| Student owns phone | 0.12 | 0.29 | 0.035 | 249 |
| F-test |  |  | 0.055 | 249 |
| **Self-reported data – week 4** |  |  |  |  |
| Gender | 0.48 | 0.42 | 0.332 | 295 |
| Grade | 1.81 | 1.79 | 0.841 | 295 |
| Student owns phone | 0.29 | 0.25 | 0.486 | 295 |
| F-test | | | 0.629 | 295 |
| **Self-reported data – week 5** | | | | |
| Gender | 0.42 | 0.46 | 0.334 | 289 |
| Grade | 1.83 | 1.66 | 0.049 | 289 |
| Student owns phone | 0.26 | 0.21 | 0.372 | 289 |
| F-test | | | 0.140 | 289 |
| **Self-reported data – week 6** | | | | |
| Gender | 0.42 | 0.49 | 0.243 | 319 |
| Grade | 1.74 | 1.8 | 0.544 | 319 |
| Student owns phone | 0.27 | 0.23 | 0.478 | 319 |
| F-test | | | 0.506 | 319 |
| **Self-reported data – week 7** | | | | |
| Gender | 0.41 | 0.37 | 0.557 | 273 |
| Grade | 1.75 | 1.90 | 0.148 | 273 |
| Student owns phone | 0.28 | 0.29 | 0.870 | 273 |
| F-test | | | 0.395 | 273 |

**Notes:** Ordinary Least Squares (OLS) regressions of student characteristics on treatment assignment for students with valid outcome data. P-values from a test of equality of means of the treatment and control groups for each covariate (jointly for the F-test), with standard errors clustered at the school level. Results indicate that student characteristics are balanced across the treatment and control groups when data are pooled together and for all weeks except week 3. For this reason, we control for student characteristics in all week-by-week regressions in Figure 2 – Panel B.

**Table S2** – Selective non-response tests for self-reported data, weeks 2 to 4

|  | Responds to SMS survey (Weeks 2 to 4) |
|---|---|
|  | (1) |
| Treatment | -0.0052 |
|  | (0.0061) |
| Control group mean | 0.051 |
| Treatment = 0 [p-value] | 0.3987 |
| Observations | 17,428 |
| R-squared | 0.0001 |

**Notes:** Ordinary Least Squares (OLS) regression with outcome variable = 1 if the student responded to the SMS surveys during the intervention (i.e. between weeks 2 and 4), and 0 otherwise. P-value from a test of equality of proportions between the treatment and control groups, with standard errors clustered at the school level. Results show that the probability of responding to SMS surveys is not systematically affected by the treatment.

**Table S3** – Selective non-response tests for self-reported data, week by week

|  | Responds to SMS survey | | | | | |
|---|---|---|---|---|---|---|
|  | Week 2 (1) | Week 3 (2) | Week 4 (3) | Week 5 (4) | Week 6 (5) | Week 7 (6) |
| Treatment | -0.0011 (0.0025) | -0.0028 (0.0022) | -0.0016 (0.0033) | -0.0016 (0.0028) | -0.0006 (0.0030) | -0.0038 (0.0027) |
| Control group mean | 0.018 | 0.017 | 0.019 | 0.018 | 0.019 | 0.019 |
| Treatment = 0 [p-value] | 0.6725 | 0.1967 | 0.6189 | 0.5628 | 0.8519 | 0.1586 |
| Observations | 16,872 | 16,828 | 16,876 | 16,872 | 16,899 | 16,852 |
| R-squared | 0.0000 | 0.0001 | 0.0000 | 0.0000 | 0.0000 | 0.0002 |

**Notes:** Ordinary Least Squares (OLS) regressions with outcome variable = 1 if the student responded to the SMS survey in each week, and 0 otherwise. P-values from tests of equality of proportions between the treatment and control groups, with standard errors clustered at the school level. Results show that the probability of responding to SMS surveys is not systematically affected by the treatment at any week.

**Table S4** – Treatment effects on student dropouts during school shutdown, for alternative definitions of the outcome variable

|  | No attendance over the last X days | |
|---|---|---|
|  | 5 days | 15 days |
|  | (1) | (2) |
| Treatment | -0.0704*** | -0.0455*** |
|  | (0.0256) | (0.0166) |
| Control group mean | 0.096 | 0.058 |
| Treatment = 0 [p-value] | 0.0074 | 0.0074 |
| Observations | 963,985 | 963,985 |
| R-squared | 0.0059 | 0.0040 |

**Notes:** ITT estimates from Ordinary Least Squares (OLS) regressions with dropouts = 1 if a student had no attendance on record over the last week (column 1) or over the last three weeks (column 2) before the winter break, and 0 otherwise. P-values from a test of equality of proportions between the treatment and control groups, with standard errors clustered at the school level. Results show that the patterns showcased by Figure 1 - Panel A are robust to alternative definitions of student dropouts during the school shutdown.

**Pre-Analysis Plan**

This randomized controlled trial was pre-registered as trial 5986 at the American Economic Association's registry for randomized controlled trial (AEA RCT Registry). The uploaded plan is detailed below.

# Does Nudging Students Decrease Learning Deficits and Dropouts During and After a Pandemic? Experimental Evidence from Covid-19 Responses in Brazil

Pre-analysis Plan

*The covid-19 pandemic has forced 1.5 billion schoolchildren in 160 countries to stay at home while schools were shut down on sanitary grounds. While several distance learning tools have been put in place in developing countries, a variety of factors raise critical concerns about learning deficits and school dropouts when schools are back, particularly amongst the most vulnerable students. This paper investigates whether sending reminders and encouragement messages to high-school students in Brazil during the pandemic increases attendance and assignment completion when it comes to distance learning, and decreases grade repetition and dropout rates in the aftermath.*

## I.     Introduction

The covid-19 pandemic has forced 1.5 billion schoolchildren in 160 countries to stay at home while schools were shut down on sanitary grounds. Brazil is no exception. The nationwide decision to shut down schools for almost the entirety of the 2020 school year in order to limit the spread of the covid-19 pandemic has forced all schools to switch to distance learning. Such rapid transition, combined with a mismatch between delivery channels and access conditions – as several State Secretariats of Education switched to online, while nearly 70 million households have no or only precarious access to internet –, are expected to severely impact learning, and potentially lead to a spike in school dropouts (Brookings, 2020; World Bank, 2020).

Schools have been trying to keep contact with their students by sending personal letters via post or by creating an online platform with tools that students can use. However, the attendance of the students, whether on the platform with the online tools or at school to pick up printed class material, is reported to be remarkably low. São Paulo State has reported that only 50% of its 3.5 million students are accessing the online learning platform daily as expected.[*]

---

[*] The State Secretariat also broadcasts content on television. It is much harder to gather data on the share of students following classes on this format daily.

With the goal of increasing engagement in distance learning – and, particularly, online attendance and assignment completion – during the pandemic, as well as limiting its effects on learning gaps and school dropouts once schools are back, the Goiás State Secretariat of Education is testing various strategies in partnership with Instituto Sonho Grande.[†] As part of those strategies, they are interested in evaluating nudges (reminders and encouragement messages) sent twice a week to high school students, directly on their mobile phones via text messages (SMS). Towards that goal, they have hired Eduq+, an educational nudgebot that has been shown to improve educational outcomes (during normal times) in Brazil and Ivory Coast.

Eduq+ nudges users twice a week with motivating facts and suggested activities to engage them in the daily school life. It also allows schools to broadcast messages to all users weekly. The intervention has been evaluated in the context of regular schooling, targeted at parents of primary school children. The nudgebot has been shown to promote large impacts on school attendance, test scores and grade promotion rates (Bettinger et al., 2020), and to decrease school dropouts by 50% across multiple primary grades (Lichand and Wolf, 2020).

The version of Eduq+ to be evaluated in this study is, however, different from that in those studies, since nudges will be sent directly to students themselves.[‡] Moreover, the context of distance learning is also much more challenging. Whether the intervention is still able to improve educational outcomes under those conditions is an empirical question.

This pre-analysis plan summarizes the design of a field experiment to test the following primary hypotheses:

1. Does nudging students increase usage of online learning tools by high school students?

   - Hypothesis: SMS nudges increase the share of students who access the online platform daily, and the share of students who hand in assignments (online or not).

2. Does nudging students mitigate the negative effects of the school shutdown on learning outcomes?

   - Hypothesis: SMS nudges improve attendance and grades, and decrease grade repetition and dropouts once face-to-face classes resume.

## II.    Intervention and experimental design

---

[†] Goiás a relatively poor state located in the Center-West region of Brazil. Instituo Sonho Grande is a non-profit organization committed to improving high-school educational outcomes in Brazilian public school.
[‡] In case they do not have their own phone, messages will be sent to the mobile phone of their primary caregivers.

The intervention has been designed by Instituto Sonho Grande and the Goiás State Secretariat of Education, with the help of Movva (the implementing partner that powers Eduq+).[§] It will take place during the months of June and July/2020, when public high schools will be randomly assigned to have their students receive two messages per week from Eduq+. 57 schools have been assigned to the treatment group, and 30 to the control group (which receives no intervention). Randomization is stratified by gender, grade and phone ownership. In case the student does not own a phone, messages will be sent to the mobile phone of his/her primary caregiver. The intervention is scheduled to be rolled out on June 9th.

Table 1: Randomization strategy - Treatment vs. Control

| Treatment | Control |
| --- | --- |
| 57 schools | 30 schools |
| 12,056 students | 6,200 students |

Table 1 above summarizes the randomization strategy for the first phase of intervention. Within the sample of 12,056 students assigned to receive nudges, less than half (5,188) own their own mobile phone and will receive messages directly. It is also important to note that not all students in the sample have access to the internet and that those who do not can pick up the printed class material once every week and hand in assignments the following week. For the purpose of this study however, we will be able to measure their outcomes in different ways.

At the end of July, we will be able estimate treatment effects on access to the online platform, and assignment completion, from administrative data provided by the Secretariat. Concretely, we have requested weekly student-level data on log in activity – or face-to-face pick-up of class materials – as well as assignment completion (again, online or offline). For those with online access, we hope to get access to daily data, which would allow us to also estimate high-frequency treatment effects through event studies. Last, after face-to-face classes resume, we will have access to administrative records on student-level attendance, grades, grade repetition and enrollment status.

The interpretation of these long-term effects will vary depending on the choice made by the Education Secretariat to continue or not the intervention after short-term results are made available. Depending on the short-term impacts of the nudges, the Education Secretariat might decide to keep testing Eduq+ for a longer period, to scale it up or to scale it down. As such, three scenarios can emerge after the first phase of the intervention: (1) the intervention continues for a longer period, keeping the treatment assignment fixed; (2) the control group starts receiving the nudges; or (3) the treatment group stops receiving the nudges. In case (1), long-term effects will reflect a combination of nudges sent during and after the school shutdown; in case (2), long-term effects will only reflect differences in the intensity of the treatment; and in case (3), long-term effects will capture persistence of treatment effects (if any).

---

§ One of the authors (Guilherme) is a co-founder and chairman at Movva (http://movva.tech).

With the number of schools and the number of students presented in Table 1, and assuming an intra-cluster correlation of 0.16 (SARESP, 2014a, 2014b) as well as conservative variance estimation for binary outcomes (assuming that 50% of students access the online platform and hand in assignments, in the control group)-, we could detect treatment effects of at least 0.8 percentage points on those outcomes[**]. Since the typical treatment effect of nudges on binary decisions is 1.7 percentage points (Dellavigna and Linos, 2020), we conclude that the design is well powered to detect relevant short-term effect sizes.

### III. Outcomes

We will document the effects of the treatments on the following categories of outcomes for students enrolled in high school (age 15 to 18):

A. Short-term outcomes: probability of logging into the online platform or picking up the material in school, probability of handed in of assignments, as measured by administrative records;

B. Long-term outcomes: attendance, grades, probability of grade repetition and probability of dropout, as measured by administrative records.

Since some students will receive messages on their own mobile phones, while for others it is their caregivers who will be nudged by Eduq+, we will estimate treatment effects within those two subgroups. Power calculations indicate that we could detect treatment effects of at least 1 and 0.9 percentage point for these two subsamples, respectively.

Since there are siblings in the data, we will remove from the main analysis cases when not all siblings are assigned to the same treatment conditions. Depending on how many siblings there are, we also plan to estimate within-family's externalities of the nudges, taking advantage of that sub-sample.

### IV. Empirical analysis

---

[**] These power calculations have been computed by clustering at the school level.

Since the intervention is randomly assigned, comparing treatment and control groups yields treatment effects of the SMS nudges on the outcomes of interest (Section III). Using ordinary least squares regressions, we will estimate:

$$Y_{smi}^j = \beta_0 + \beta_1 T_{sm} + \theta_s + u_{smi}$$

Where:

- $Y_{smi}^j$: Outcome variable j for student i at school m and stratum s;
- $T_m$: Indicator variable equal to 1 if students I in school m and stratum s is assigned to receive SMS nudges, 0 otherwise;
- $\theta_s$: stratum fixed effects.

We cluster standard errors at the school level, since that is the level at which the intervention is randomly assigned. We are interested in testing $\beta_1 = 0$.